\documentclass{article}
\usepackage{amsmath,amssymb,graphicx,microtype,hyperref}

\def\be{\begin{eqnarray}}
\def\ee{\end{eqnarray}}
\def\nn{\nonumber}

\def\l[{\phantom.[}

\hoffset= - 1.0in         

\begin{document}

\title{\vspace{1cm}{\Large {\bf On elementary proof of AGT relations from six dimensions
}\vspace{.2cm}}
\author{{\bf A. Mironov$^{a,b,c,d,}$}\footnote{mironov@lpi.ru; mironov@itep.ru}, \ {\bf A. Morozov$^{b,c,d,}$}\thanks{morozov@itep.ru}, \ and \ {\bf Y. Zenkevich$^{b,d,e,}$}\thanks{yegor.zenkevich@gmail.com}}
\date{ }
}

\maketitle

\vspace{-6cm}

\begin{center}
\hfill FIAN/TD-13/15\\
\hfill IITP/TH-20/15\\
\hfill ITEP/TH-32/15\\
\hfill INR-TH/2015-037
\end{center}

\vspace{3.2cm}

\begin{center}
$^a$ {\small {\it Lebedev Physics Institute, Moscow 119991, Russia}}\\
$^b$ {\small {\it ITEP, Moscow 117218, Russia}}\\
$^c$ {\small {\it Institute for Information Transmission Problems, Moscow 127994, Russia}}\\
$^d$ {\small {\it National Research Nuclear University MEPhI, Moscow 115409, Russia }}\\
$^e$ {\small {\it Institute of Nuclear Research, Moscow 117312, Russia }}
\end{center}

\vspace{1cm}

\begin{abstract}
The actual definition of the Nekrasov functions participating in the AGT relations implies a peculiar choice of contours in the LMNS and Dotsenko-Fateev integrals. Once made explicit and applied to the original triply-deformed (6-dimensional) version of these integrals, this approach reduces the AGT relations to symmetry in $q_{1,2,3}$, which is just an elementary identity for an appropriate choice of the integration contour (which is, however, a little non-traditional). We illustrate this idea with the simplest example of ${\cal N}=(1,1)$ $U(1)$ SYM in six dimensions, however all other cases can be evidently considered in a completely similar way.
  \end{abstract}

\vspace{.5cm}

\bigskip

\bigskip

\bigskip

Seiberg-Witten theory and its quantization \cite{SW}--\cite{GM},
provided by Nekrasov's evaluation \cite{Nek} of the LMNS integrals
\cite{LMNS}, is one of the cornerstones of modern theoretical physics.
In different ways this story is related to a majority of other important
subjects.  In particular, the AGT relations \cite{AGT}
provide a connection to $2d$ conformal theories \cite{CFT,DF} and, perhaps,
further to the generic stringy AdS/CFT correspondence. Lifting the original
four-dimensional story to five and six dimensions makes a contact with $q$-
and elliptic Virasoro algebras~\cite{qellVir}, with (refined)
topological string theories~\cite{ref} and, finally, with
still mysterious double-elliptic integrable systems \cite{dell}. As
usual, things are rather obscure in low dimensions and get clarified
when their number increases, at expense of an undeveloped language to
describe these simple, but somewhat non-classical structures. In this
letter, we provide a brief summary of the last years efforts to
understand and prove refinement procedures on one side and the AGT
relations on another, and emphasize that the choice of the right
language is sufficient to convert the latter into an elementary
identity. Below is a very brief, though exhaustive presentation. A less formal and more
traditional version will be provided in a longer accompanying paper.

\bigskip

We consider the integrals
\begin{equation}
Z_\gamma\{Q,q\} = \int_\gamma dx^N F\{x_i|Q,q\}
\label{Integral}
\end{equation}
where $F$ are basically the products and ratios (perhaps, infinite) of
Van-der-Monde like quantities $\prod_{i\neq j}(x_i- c x_j)$ over some
$Q,q$-dependent families $C$ of parameters $c$:
\begin{equation}
F\{x_i|Q,q\} \sim \frac{\prod_{c\in C^+}  \prod_{i\neq j}(x_i- c x_j)}
{\prod_{c\in C^-}  \prod_{i\neq j}(x_i- c x_j)}
\label{integrandVdm}
\end{equation}
Such $F$ has a variety of poles.
The integration contour $\gamma$ can be chosen to pick up some of these poles,
so that the integral becomes a sum of residues over them.
In other words, $\gamma$ defines a set $\Pi_\gamma$ of poles,
\begin{equation}
x_i = x_i^\pi(Q,q) \ \ \ \text{for}\ \  \ \pi\in\Pi_\gamma
\end{equation}
so that
\begin{equation}
\boxed{
Z_\gamma\{C_\pm\} = \sum_{\pi \in\Pi_\gamma} F^\prime_\pi\{Q,q\}
}
\label{Sum}
\end{equation}
where
\begin{equation}
F^\prime_\pi\{Q,q\} = \left.F\{x_i|Q,q\}\prod_{i=1}^N \Big(x_i-x_i^\pi(Q,q)\Big)\right|_{x_i = x_i^\pi(Q,q)}
\end{equation}
and we switched from $Q,q$ to notation with $C_\pm$, which is sometime more adequate.

Integrand (\ref{integrandVdm}) can be rewritten in terms of time-variables
$p_n = \sum_ix_i^n$,
\begin{equation}
F\{x_i|Q,q\} \approx \exp \left(\mp \sum_{c\in C_\pm}  \frac{c^n p_np_{-n}}{n}\right)
\label{integrandtimes}
\end{equation}
which can be further expanded into the Schur/Macdonald polynomials of various types \cite{SM,Zenk},
depending on particular sets $C_\pm$.

This technique was intensively used to demonstrate that the substitution
(\ref{Integral}) $\longrightarrow$ (\ref{Sum}) is the outcome of more standard
procedures like Selberg integration, and actually (\ref{Sum}),
perhaps, in the form
\begin{equation}
\boxed{
Z_\gamma\{C_\pm\} = \sum_{\pi \in\Pi_\gamma} \exp \left(\mp \sum_{c\in C_\pm}
\frac{c^n p_n^\pi p_{-n}^\pi}{n}\right)
}
\label{Sump}
\end{equation}
can be taken as a {\it definition} of relevant quantities in Seiberg-Witten-Nekrasov theory.
Moreover, the sets $\Pi_\gamma$ in this context are {\it postulated} to be some
collections of ordinary or 3d Young diagrams.
The role of parameters $Q$ and $q$ is different in these theories:
$Q$'s are {\it moduli} (dimensions, brane lengths, masses, couplings depending on the
preferred language), while $q$'s are theory parameters
(like the compactification radii of the 5-th, 6-th and 11-th dimensions).
Technically, $Q$ and $q$ enter in different ways into the sets $C_\pm$ and $\Pi_\gamma$.

The AGT relations are then identities between sums/integrals with
different sets $\{C_\pm\}$, describing the LMNS and Dotsenko-Fateev
integrals. The identities can be actually understood as a symmetry in
parameters $q$, which is, however, obvious only when their number is
at least three, $q=\{q_1,q_2,q_3\}$, while it gets obscure in the
limits when some of these parameters turn to zero (which corresponds
to reducing the dimension of associated Yang-Mills theory from six to five
and four). Thus, understanding the ``$6$-dimensional and $M$-theory'' origin
of the theory allows one to provide an {\bf elementary proof of the AGT relations}. Now we provide a simple illustration of this thesis
with quite a non-trivial (!) example of the AGT identity.

\bigskip

The LMNS integral in $5d\ $ $U(1)$ theory with fundamental
matter hypermultiplets in a specific point of the moduli space, which will be our basic example is (the integral appeared, e.g., in \cite{CNO})
\begin{equation}
Z_{LMNS}=\int d^Nx \prod_{i\neq j} \frac{(x_i-x_j)(x_i-t\tilde t x_j)}{(x_i-tx_j)(x_i-\tilde t x_j)}
\end{equation}
The Dotsenko-Fateev (DF) like integral \cite{DFl1,DFl2} describing the AGT-related conformal block of
the $q$-deformed Virasoro algebra is~\cite{MMSha,AS},
\begin{equation}
Z_{DF} = \int d^Nx \prod_{k \geq 0}\prod_{i\neq j} \frac{x_i - q^k
  x_j}{x_i - t q^k x_j}
\end{equation}
It is clear that these integrals are two different limits of the
following ``affine Selberg integral''
\begin{equation}
  Z(q,t,\tilde t) = \int  d^Nx \prod_{i\neq j} \prod_{k=0}^\infty
  \frac{(x_i-q^kx_j)(x_i-t\tilde t q^k x_j)}{(x_i-tq^k x_j)(x_i-\tilde t q^k x_j)}\label{pvsE}
\end{equation}
namely
\begin{equation}
Z_{LMNS}(t,\tilde t) = Z(q=0,t,\tilde t)
\end{equation}
and
\begin{equation}
Z_{DF}(q,t) = Z(q,t, \tilde t=0)
\end{equation}
Our claim is that the AGT relation
\begin{equation}
\boxed{
Z_{LMNS}(q,t) = Z_{DF}(q,t)
}
\label{AGT}
\end{equation}
is just a trivial corollary of the symmetry
\begin{equation}
\boxed{
Z(q,t,\tilde t)= Z(\tilde t,t,q)=\text{four other permutations}
}
\label{symmZ}
\end{equation}

As to (\ref{symmZ}), it is, indeed, an elementary identity,
provided this integral is {\it defined} as (\ref{Sump}),\footnote{Since one can not just evaluate the integrand of (\ref{pvsE}) at the poles, we actually compute the ratio of the integrands at the pole and at the pole corresponding to the empty diagram (this ratio is finite). The same trick was used in \cite{AS} and leads to an inessential normalization factor.}
\begin{equation}
Z = \sum_{\pi \in \Pi} \exp \left(-\sum_{n\geq 1}
\frac{(1-t^n)(1-\tilde t^n)}{1-q^n}\frac{p_n^\pi p_{-n}^\pi}{n}
\right)
\end{equation}
with $N=\infty$ and $\Pi$ being the set of all 3d partitions, and
\begin{equation}
\{x_i^\pi\} = \{q^{\pi_{b,c}-1}\cdot t^{b-1}\cdot \tilde t^{c-1} \}
\label{setPiAGT}
\end{equation}
with $\pi_{b,c}$ being the height of partition
$\pi\in \Pi$ at the point $b,c$ (see for a similar calculation \cite{NOM}).
Indeed, then
\begin{multline}
p_n^\pi = \sum_{b,c\geq 1} q^{n(\pi_{b,c}-1)}t^{\,n(b-1)}\tilde t^{\,n(c-1)}
=\sum_{b,c\geq 1} t^{\,n(b-1)}\tilde t^{\,n(c-1)}\Big(1+(q^{n(\pi_{b,c}-1)}-1\Big)= \nn \\
= \frac{1}{(1-t^n)(1-\tilde t^n)} - (1-q^n)\sum_{(a,b,c)\in\pi}q^{n(a-1)}t^{\,n(b-1)}\tilde t^{\,n(c-1)}
= \frac{E_\pi(q^n,t^n,\tilde t^n)}{(1-t^n)(1-\tilde t^n)}
\label{pvsE}
\end{multline}
The function
\begin{equation}
E_\pi(q_1,q_2,q_3)=1-(1-q_1)(1-q_2)(1-q_3)\cdot
\!\!\!\!
\sum_{(a,b,c)\in\pi}q_1^{a-1}q_2^{b-1}q_3^{c-1}
\end{equation}
itself is not symmetric in $q_{1,2,3}$, but permutations of variables transpose the
3d Young diagram $\pi$, so that the sum over all $\pi$ in
\begin{equation}
\boxed{
Z(q,t,\tilde t) = \sum_{\text{all 3d partitions\ }\pi } \exp \left(-\sum_{n\geq 1}
\frac{E_\pi(q^n,t^n,\tilde t^n)E_\pi(q^{-n},t^{-n},\tilde t^{-n})}
{n(1-q^n)(1-t^n)(1-\tilde t^n)}\right)
}
\label{symoZ}
\end{equation}
is symmetric, i.e. (\ref{symmZ}) is true. Notice also that this
expression exactly reproduces the instanton partition function of the
$\mathcal{N} = (1,1)$ $U(1)$ gauge theory in six
dimensions~\cite{N11}.

One can check that the choice~(\ref{setPiAGT}) is consistent with the
usual definitions of $Z_{LMNS}$ and $Z_{AGT}$: the residues of these
two integrals are enumerated by ordinary partitions, which are just
two different projections of a single 3d partition.
The integral~(\ref{setPiAGT}) easily generalizes to $U(N)$ theory,
which basically amounts to considering a \emph{set} of 3d Young
diagrams instead of one. One can also introduce external parameters
$Q$, corresponding to the massive matter of the gauge theory, and to the vertex
operators in CFT respectively, but we postpone all these
details, including a more accurate explanation of~(\ref{pvsE}) to a
longer paper, to avoid unnecessary complications in the presentation
of a simple idea.
Let us notice that
the AGT relation, which we consider here is actually a combination of
the standard AGT relation and the spectral duality of conformal block \cite{SD,Zenk,AS},
so that e.g. the $U(N)$ gauge theory corresponds to the $N$-point conformal
block.

\bigskip

To conclude, we claim that after an appropriate extension and with clever
definitions like (\ref{Sump}) the somewhat {\it transcendental} AGT
relation~(\ref{AGT}) between the LMNS and DF-like integrals reduces to
{\it elementary symmetry properties} (\ref{symmZ}) of sums over 3d partition
like (\ref{symoZ}).  This symmetry seems to be the true meaning
of the Hubbard-Stratonovich duality of \cite{MMSha,MMSha1}.

\section*{Acknowledgements}

Our work is partly supported by grants NSh-1500.2014.2 (AM's) and 15-31-20832-Mol-a-ved (A.Mor.), by RFBR grants 16-01-00291 (A.Mir.) and 16-02-01021 (A.Mor. and Y.Z.), by joint grants 15-51-50034-YaF, 15-51-52031-NSC-a, by 14-01-92691-Ind-a, by the Brazilian National Counsel of Scientific and Technological Development (A.Mor.).

\end{document}